\documentclass[preprint,12pt]{elsarticle}

\usepackage{amssymb}
\usepackage{amsmath}

\usepackage{pdfpages}
\usepackage{threeparttable}
\usepackage{tabularx}
\usepackage{float}
\usepackage{ulem}
\usepackage{dcolumn}
\usepackage{siunitx}
\usepackage{makecell}
\usepackage{booktabs}
\usepackage{graphicx} 
\usepackage{tikz}

\usepackage{algorithm}
\usepackage{algpseudocode}
\usepackage{multirow}
\usepackage{caption}
\usepackage{geometry}
\usepackage[colorlinks=true,linkcolor=blue,citecolor=blue,urlcolor=blue]{hyperref}

\journal{Computational Materials Science}

\begin{document}

\begin{frontmatter}



\title{Short-Range Order Based Ultra Fast Large-Scale Modeling of High-Entropy Alloys}


\author[1]{Caimei Niu}
\ead{caimeiniu@stu.pku.edu.cn}

\author[2]{Lifeng Liu \corref{cor1}}
\ead{lfliu@pku.edu.cn}

\affiliation[1]{organization={School of Software and Microelectronics, Peking University}, 
            city={Beijing},
            postcode={100871}, 
            country={China}}
            
\affiliation[2]{organization={School of Integrated Circuits, Peking University}, 
            city={Beijing},
            postcode={100871}, 
            country={China}}

\cortext[cor1]{Corresponding author}

\begin{abstract}

High-Entropy Alloys (HEAs) exhibit complex atomic interactions, with short-range order (SRO) playing a critical role in determining their properties. Traditional methods, such as Monte Carlo generator of Special Quasirandom Structures within the Alloy Theoretic Automated Toolkit (ATAT-mcsqs), Super-Cell Random APproximates (SCRAPs), and hybrid Monte Carlo-Molecular Dynamics (MC-MD)—are often hindered by limited system sizes and high computational costs. In response, we introduce PyHEA, a Python-based toolkit with a high-performance C++ core that leverages global and local search algorithms, incremental SRO computations, and GPU acceleration for unprecedented efficiency. When constructing random HEAs, PyHEA achieves speedups exceeding 333,000× and 13,900× over ATAT-mcsqs and SCRAPs, respectively, while maintaining high accuracy. PyHEA also offers a flexible workflow that allows users to incorporate target SRO values from external simulations (e.g., LAMMPS or density functional theory (DFT)), thereby enabling more realistic and customizable HEA models. As a proof of concept, PyHEA successfully replicated literature results for a 256,000-atom Fe–Mn–Cr–Co system within minutes—an order-of-magnitude improvement over hybrid MC–MD approaches. This dramatic acceleration opens new possibilities for bridging theoretical insights and practical applications, paving the way for the efficient design of next-generation HEAs.

\end{abstract}






\begin{highlights}
\item High Performance: PyHEA delivers unprecedented speedups of 333,000× over ATAT-mcsqs and 13,900× over SCRAPs in random structure generation, achieving superior efficiency without compromising accuracy.

\item Large-Scale Compatibility: PyHEA models systems with over 256,000 atoms in seconds, enabling simulations at scales relevant to experimental studies and unlocking new opportunities for high-entropy alloy research.

\item GPU Acceleration: PyHEA’s CUDA-based parallel Monte Carlo implementation provides over 1,000× speedup compared to CPU-based calculations, dramatically reducing simulation times for large-scale configurations.

\item User-Friendly Python Ecosystem: Built on a Python interface with a high-performance C++ core, PyHEA provides an accessible yet powerful solution for structure generation, SRO analysis, and data visualization, with seamless integration into existing workflows.

\end{highlights}


\end{frontmatter}



\section{Introduction}
\label{Intro}
High-Entropy Alloys (HEAs), introduced by Jien-Wei Yeh et al. in 2004 \cite{yeh2004nanostructured}, have emerged as a transformative class in materials science. Characterized by the presence of five or more principal elements in quasi-equimolar ratios (typically 3-55\% atomic percent each), HEAs exhibit a unique combination of remarkable properties, including high strength, hardness, and exceptional resistance to oxidation and corrosion \cite{miracle2017critical, george2019high, li2021mechanical, yeh2007high}. These attributes position HEAs as promising materials for applications in extreme environments, such as aerospace, nuclear, and energy sectors. Furthermore, the unconventional nature of HEAs challenges traditional alloy design theories, which spurred significant research interest in their underlying principles \cite{zhang2014microstructures, li2016metastable}.

Recent advancements have highlighted the critical role of microstructure in determining HEAs’ extraordinary properties, particularly the influence of short-range order (SRO) \cite{zhang2012alloy, cowley1950approximate, yin2020ab, cohen1962some, zhao2021local}. 
SRO refers to the local atomic arrangement within an alloy, a phenomenon that becomes increasingly prominent in complex concentrated alloys, even under rapid solidification conditions \cite{ding2019tuning, han2024ubiquitous}. The impact of SRO on mechanical performance and thermal stability is profound, influencing elastic modulus, stiffness, and ductility by optimizing the balance between strength and toughness \cite{pickering2016high, chen2021simultaneously, liu2024atomistic}. These findings underscore the potential for leveraging SRO control to tailor HEA properties, providing a theoretical framework for the development of new high-performance materials.

The primary challenge in designing HEAs with precise SRO control lies in the vast configurational space. For an equiatomic HEA with $N$ total atoms and $M$ different principal elements, the configurational space size $\Omega$ is given by:
\begin{equation}
\Omega = \frac{N!}{\left(\frac{N}{M}!\right)^M}.
\end{equation}
This expression reveals the astronomical complexity of the problem—even a modest Cantor alloy (CrMnFeCoNi) with just 250 atoms yields approximately $10^{170}$ possible configurations \cite{ferrari2023simulating, zhang2022calphad}. Moreover, realistic applications demand simulations of tens or even hundreds of thousands of atoms to fully capture mechanical and structural properties \cite{seol2022mechanically}, further amplifying this complexity.

Current approaches to HEA modeling face significant limitations in addressing this challenge. Static structure construction methods like the Monte Carlo generator of Special Quasirandom Structures within the Alloy Theoretic Automated Toolkit (ATAT-mcsqs) \cite{ghosh2008first, van2013efficient, van2019alloy} and Super-Cell Random APproximates (SCRAPs) \cite{singh2021accelerating} rely on Special Quasirandom Structures (SQS) and hybrid Monte Carlo approaches to generate random alloy configurations respectively. While these methods are effective for handling relatively small supercells (tens to hundreds of atoms), they become computationally prohibitive when dealing with larger systems. \cite{he2024quantifying}

In parallel, Hybrid Monte Carlo-Molecular Dynamics (MC-MD) methods offer another alternative approach by incorporating MC steps into extended MD simulations. Although hybrid MC-MD can theoretically handle systems with hundreds of thousands of atoms \cite{salloom2022atomic, ferrari2023simulating, liu2023machine}, its practical effectiveness depends heavily on the availability of highly accurate interatomic potentials. Even when such potentials exist, generating configurations that reliably reflect SRO demands substantial computational resources, often requiring hours or days of calculation time.

To address these challenges, we introduce PyHEA, a high-performance computational framework that revolutionizes large-scale HEA modeling through three key innovations:
\begin{itemize}
    \item \textbf{Enhanced efficiency:} 
    We benchmarked PyHEA against ATAT-mcsqs and SCRAPs by generating a series of random structures. As summarized in \autoref{tab:performance}, PyHEA achieves speedups of up to $3.33\times10^5$ over ATAT-mcsqs and $1.39\times10^4$ over SCRAPs, all without sacrificing the accuracy of the generated configurations. These results highlight PyHEA’s robust numerical stability and superior modeling efficiency, establishing it as a powerful alternative to existing methods.
    
    \item \textbf{Unprecedented scalability:} 
    Beyond random structures, PyHEA allows users to specify target SRO values derived from experiments or prior simulations, enabling the modeling of temperature-dependent ordering and other specialized scenarios. When such SRO data are unavailable or difficult to obtain, as introduced in ~\autoref{sec:application}, we propose an integrated workflow that employs smaller-scale \textit{ab initio} (DFT) or LAMMPS simulations to compute the desired SRO. PyHEA then leverages these inputs to construct large-scale alloys, effectively streamlining conventional MC–MD processes. For instance, in modeling a 256,000-atom Fe–Mn–Cr–Co system, PyHEA provides a 1000× speedup in configuration generation. Even accounting for the time required to prepare target SRO values, an overall improvement of at least an order of magnitude is retained, making large-scale HEA simulations far more tractable.

    \item \textbf{User-friendly software ecosystem:} Built on Python with a high performance C++ core, PyHEA combines accessibility with computational efficiency. Released under the LGPL-3.0 license, it provides an end-to-end solution including structure generation, SRO analysis, and data visualization. The modular design facilitates easy integration with existing workflows and enables rapid customization for specific research needs.

\end{itemize}
In the following sections, we detail PyHEA's methodology, demonstrate its performance advantages through comprehensive benchmarks, and showcase its practical applications in a large scale HEA system. Our results establish PyHEA as a transformative tool that bridges the gap between theoretical understanding and practical simulation of large-scale HEA systems, opening new possibilities for materials design and optimization.

\section{Results}
\label{sec:results}
\subsection{Parameter Optimization Analysis}
\label{sec:param_opt}

\begin{figure}[!htb]
    \centering
    \includegraphics[width=\textwidth]{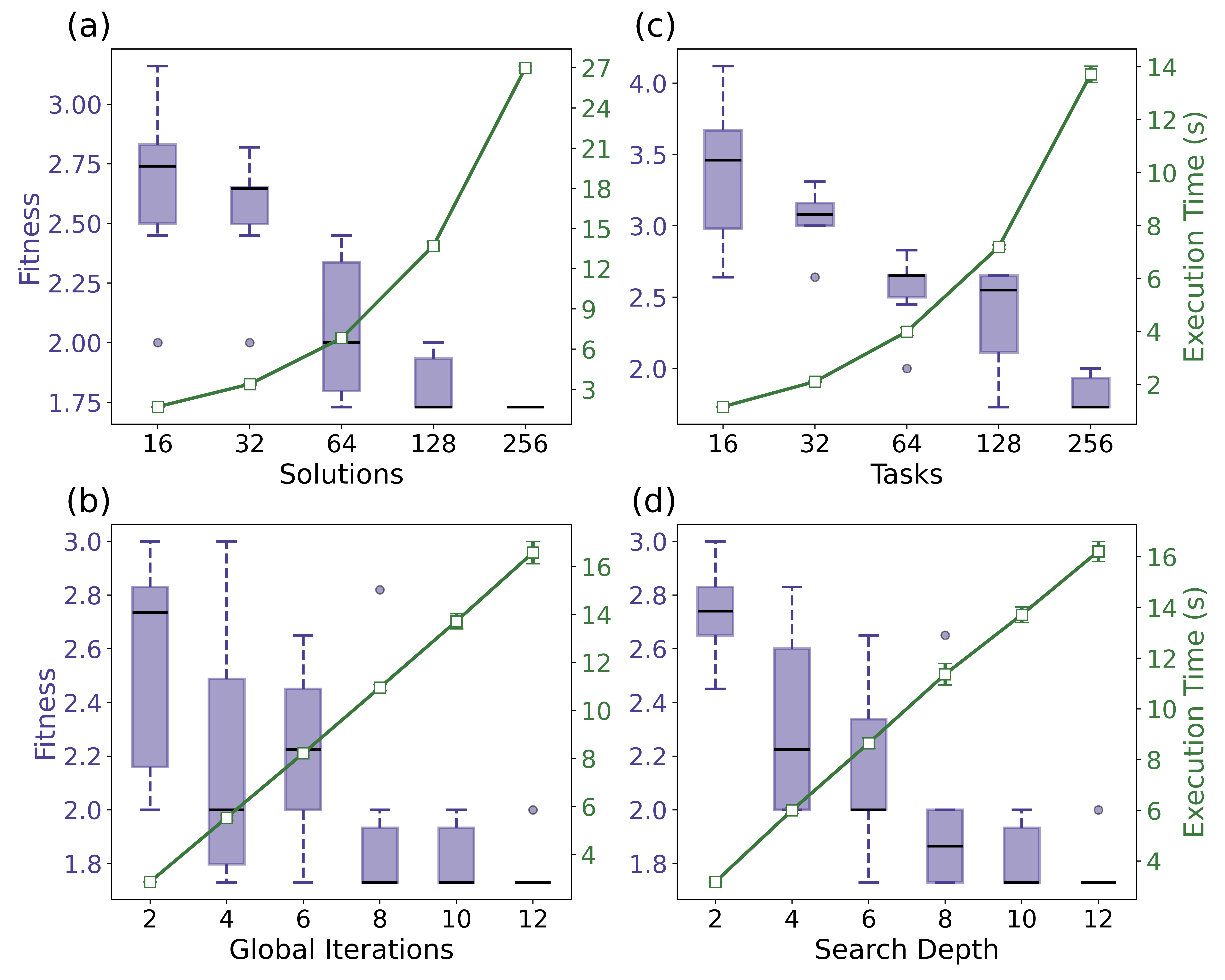}
    \caption{
    Influence of input parameters on fitness (left axis) and execution time (right axis):
    (a) Number of solutions, (b) global iterations, (c) number of tasks, and (d) search depth.
    }
    \label{fig:setting}
\end{figure}

In this subsection, we examine how four key parameters influence both the optimization quality and computational efficiency when constructing atomic configurations for HEAs:
1)~\textit{Number of solutions}, 
2)~\textit{Global iterations}, 
3)~\textit{Number of tasks}, and
4)~\textit{Search depth}.
These parameters govern the search algorithm outlined in \autoref{sec:global_search} and \autoref{sec:mc}, while the fitness function used to evaluate candidate configurations is defined in \autoref{sec:fitness_calculation} (see \autoref{eq:fitness_function}).

Conceptually, the fitness function serves as a “loss function” in an optimization context: it quantifies how “far” a candidate atomic configuration is from the target SRO. Lower fitness values indicate closer alignment with the desired SRO. By iteratively minimizing this fitness, the algorithm refines candidate configurations to achieve the prescribed atomic ordering.

The first parameter, \textit{number of solutions}, refers to the total number of candidate configurations (atomic arrangements) evaluated in parallel during each global search iteration. As illustrated in \autoref{fig:setting}(a), increasing the number of solutions from a smaller baseline up to around 64 significantly reduces the fitness, indicating better alignment with the target SRO. Beyond 64 solutions, additional gains in accuracy become negligible, while execution time increases sharply. Therefore, a range of 64--128 solutions strikes a balance between accuracy and computational efficiency.

The second parameter, \textit{global iterations}, specifies the total number of optimization cycles. \autoref{fig:setting}(b) shows that fitness improves substantially up to about six iterations. Beyond this point, further improvements are marginal compared to the additional runtime. Hence, six global iterations suffice in most cases to achieve robust SRO alignment without incurring excessive computational costs.

Parallelism, governed by the \textit{number of tasks}, is the third parameter under consideration. As shown in \autoref{fig:setting}(c), increasing the number of tasks up to around 128 enhances the consistency of fitness outcomes. However, execution time grows considerably beyond 128 tasks, while fitness gains diminish. A task count between 64 and 128 is recommended to maintain an effective balance between throughput and result fidelity.

The final parameter, \textit{search depth}, determines the thoroughness of local refinements during each iteration. \autoref{fig:setting}(d) demonstrates that deeper searches (greater than a depth of 8) lead to lower fitness values, signifying more accurate SRO alignment. However, deeper searches also incur higher computational costs. Consequently, a search depth of approximately 8--10 provides an optimal trade-off between solution precision and runtime.

Building on these insights, we propose two configurations tailored to different priorities:
\begin{itemize}
    \item \textbf{Precision-Optimized Configuration}: (128 solutions, 10 global iterations, 256 tasks, search depth 10). This setup achieves the highest fidelity to the target SRO but requires longer execution times.
    \item \textbf{Performance-Optimized Configuration}: (64 solutions, 6 global iterations, 64 tasks, search depth 8). This arrangement maintains good SRO alignment while minimizing runtime, yielding a practical balance between accuracy and computational efficiency.
\end{itemize}

By choosing one of these configurations—or adjusting parameters within the suggested ranges—users can tailor the optimization process to either prioritize maximum accuracy or expedite computations, depending on the demands of their particular HEA simulations.

\subsection{Comparison with State-of-the-Art Methods}

\begin{figure}[!htb]
    \centering
    \includegraphics[width=\textwidth]{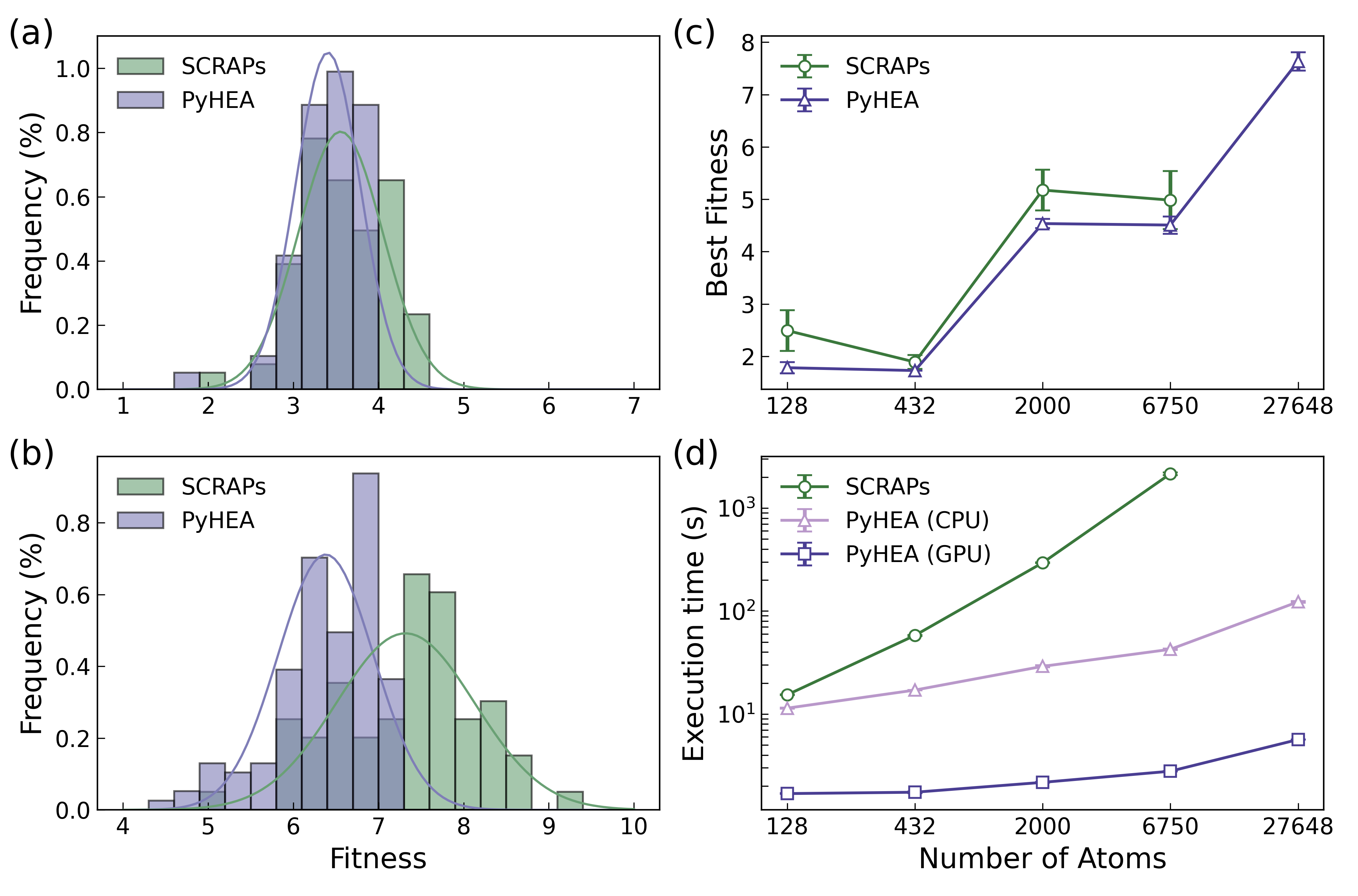}
    \caption{
    Comparison of PyHEA and SCRAPs in terms of fitness accuracy and execution time.
    (a) and (b) show the distribution of fitness values (lower values indicate higher accuracy) for systems containing 128 atoms and 2000 atoms, respectively. (c) illustrates the best fitness values as a function of system size. (d) presents the execution times for SCRAPs, PyHEA (CPU), and PyHEA (GPU) across various system sizes. Both SCRAPs and PyHEA (CPU) were tested on the same CPU cluster equipped with an Intel Xeon Gold 6132 CPU @ 2.60 GHz using 24 processors, while PyHEA (GPU) utilized an NVIDIA 4090 GPU.
    }
    \label{fig:compare}
\end{figure}

To evaluate the performance of PyHEA, we conducted a comparative analysis against SCRAPs, a state-of-the-art method in HEA modeling\cite{singh2021accelerating}. This comparison focused on evaluating both fitness quality and computational efficiency across varying system sizes.

\autoref{fig:compare}(a-b) present the fitness distributions for typical bcc configurations with 128 atoms and 2000 atoms, respectively. PyHEA consistently exhibits a more concentrated fitness distribution compared to SCRAPs, particularly as the system size increases. In the case of the 2000-atom system, PyHEA demonstrates a narrower spread with a higher peak frequency around the optimal fitness value, signifying more stable and reliable outcomes. Conversely, SCRAPs shows a broader distribution of fitness values, indicating that PyHEA's optimization strategy is more robust, especially in larger systems.

The enhanced fitness quality of PyHEA is further corroborated by \autoref{fig:compare}(c), which tracks the optimal fitness values as system size increases. While PyHEA and SCRAPs achieve comparable fitness levels for smaller systems (e.g., 128 and 432 atoms), PyHEA surpasses SCRAPs as the system size exceeds 2000 atoms, highlighting its scalability and the effectiveness of its optimization approach. This is particularly crucial for large-scale HEA simulations where fitness convergence is increasingly challenging.

A key advantage of PyHEA is its computational efficiency, particularly with GPU acceleration.\autoref{fig:compare}(d) compares the execution times of PyHEA (in both CPU and GPU implementations) and SCRAPs as a function of system size. SCRAPs faces significant scalability challenges, with execution times increasing rapidly as the system size grows. Notably, SCRAPs becomes computationally prohibitive at approximately 6750 atoms, beyond which further calculations are unfeasible.

In contrast, PyHEA’s GPU implementation seamlessly handles larger system sizes. At the largest system size of 27,648 atoms, PyHEA’s GPU-accelerated version achieves a performance improvement of four orders of magnitude compared to SCRAPs. This significant boost in computational efficiency allows PyHEA to handle system sizes that are beyond the practical capacity of SCRAPs.

To further validate PyHEA's acceleration benefits, we also conducted a comparison with another widely used MC-based tool, ATAT-mcsqs. \autoref{tab:performance} summarizes the performance metrics based on serial execution across PyHEA (CPU and GPU), SCRAPs, and ATAT-mcsqs. While SCRAPs and ATAT-mcsqs perform comparably for smaller system sizes, they both struggle as the number of atoms increases. Like SCRAPs, ATAT-mcsqs becomes impractical beyond a few hundred atoms. Conversely, PyHEA's GPU implementation maintains substantially lower execution times across all tested system sizes, underscoring its performance advantages and scalability. For a comprehensive comparison of parallel CPU performance between these tools, readers are referred to \autoref{tab:parallel} in the Appendix.

\begin{table}[!htb]
    \centering
    \begin{threeparttable}
    \caption{Performance comparison between PyHEA and existing methods (ATAT-mcsqs and SCRAPs) for optimizing atomic structures in HEAs, focusing on serial execution performance (parallel performance data available in Appendix). The table shows computation times (in minutes) and speedup ratios for both body-centered cubic (bcc) and face-centered cubic (fcc) structures across varying system sizes (\#At) and numbers of species (\#Sp). Both PyHEA and SCRAPs were executed in serial mode using a single CPU core (Intel Xeon Gold 6132 \text{@} 2.60GHz), with additional GPU (NVIDIA 4090) results shown for PyHEA. The optimizations used 128 candidate solutions, 10 global iterations, 256 parallel MC tasks, and a search depth of 10. Supercell sizes followed $S = A \cdot L^3$, where $A = 2$ for bcc or $A = 4$ for fcc structures. All calculations optimized SRO over 3 coordination shells. Dashes (--) indicate computationally infeasible cases, while speedup values are shown relative to SCRAPs[ATAT in brackets].}

    \label{tab:performance}
    \begin{tabular}{c|cc|c|cc|c|cc}
        \Xhline{0.7pt}
        \multirow{2}{*}{Type} & \multirow{2}{*}{\#Sp} & \multirow{2}{*}{\#At} & ATAT- & SCRAPs & PyHEA & PyHEA & \multicolumn{2}{c}{Speedup} \\
        & & & mcsqs & Serial & Serial & GPU & CPU & GPU \\
        \Xhline{0.7pt}
        bcc & 3 & 54 & \textgreater 1.4k\tnote{a} & 2.7 & 3.2 & .03 & .9[438] & 98[47k]\\
        bcc & 4 & 128 & \textgreater 10k\tnote{a} & 6.5 & 4.2 & .03 & 1.6[2.4k] & 233[333k]\\
        bcc & 4 & 432 & -- & 24.2 & 5.6 & .03 & 4.3 & 835\\
        bcc & 4 & 1024 & -- & 103 & 6.2 & .03 & 16.5 & 3.2k\\
        bcc & 4 & 2048 & -- & 276 & 7.0 & .04 & 39.4 & 7.9k\\
        bcc & 5 & 2048 & -- & 286 & 9.3 & .04 & 30.1 & 7.9k\\
        bcc & 5 & 6.8k & -- & 639 & 14.4 & .05 & 44.5 & 13.9k\\
        bcc & 6 & 28k & -- & -- & 39.6 & .09 & -- & --\\
        \hline
        fcc & 3 & 108 & -- & 17.4 & 3.9 & .01 & 4.5 & 3.5k\\
        fcc & 4 & 256 & -- & 68.8 & 5.5 & .01 & 12.5 & 11.5k\\
        fcc & 5 & 500 & -- & 260 & 9.4 & .01 & 27.7 & 37.2k\\
        fcc & 5 & 4k & -- & 1.6k & 14.2 & .01 & 110 & 111k\\
        fcc & 4 & 256k & -- & -- & 1.2k & .58 & -- & --\\
        \Xhline{0.7pt}
    \end{tabular}
    \begin{tablenotes}
    \small
        \item[a] Notation: \#Sp = number of elemental species.
        \item[b] Notation: \#At = total number of atoms in the system.
        \item[c] Speedup ratios: value vs SCRAPs [value vs ATAT in brackets].
        \item[d] ATAT-mcsqs data from Ref. \cite{singh2021accelerating}. All times in minutes.
    \end{tablenotes}
    \end{threeparttable}
\end{table}

\subsection{Application}
\label{sec:application}

\begin{figure}[!htb]
    \centering
    \includegraphics[width=\textwidth]{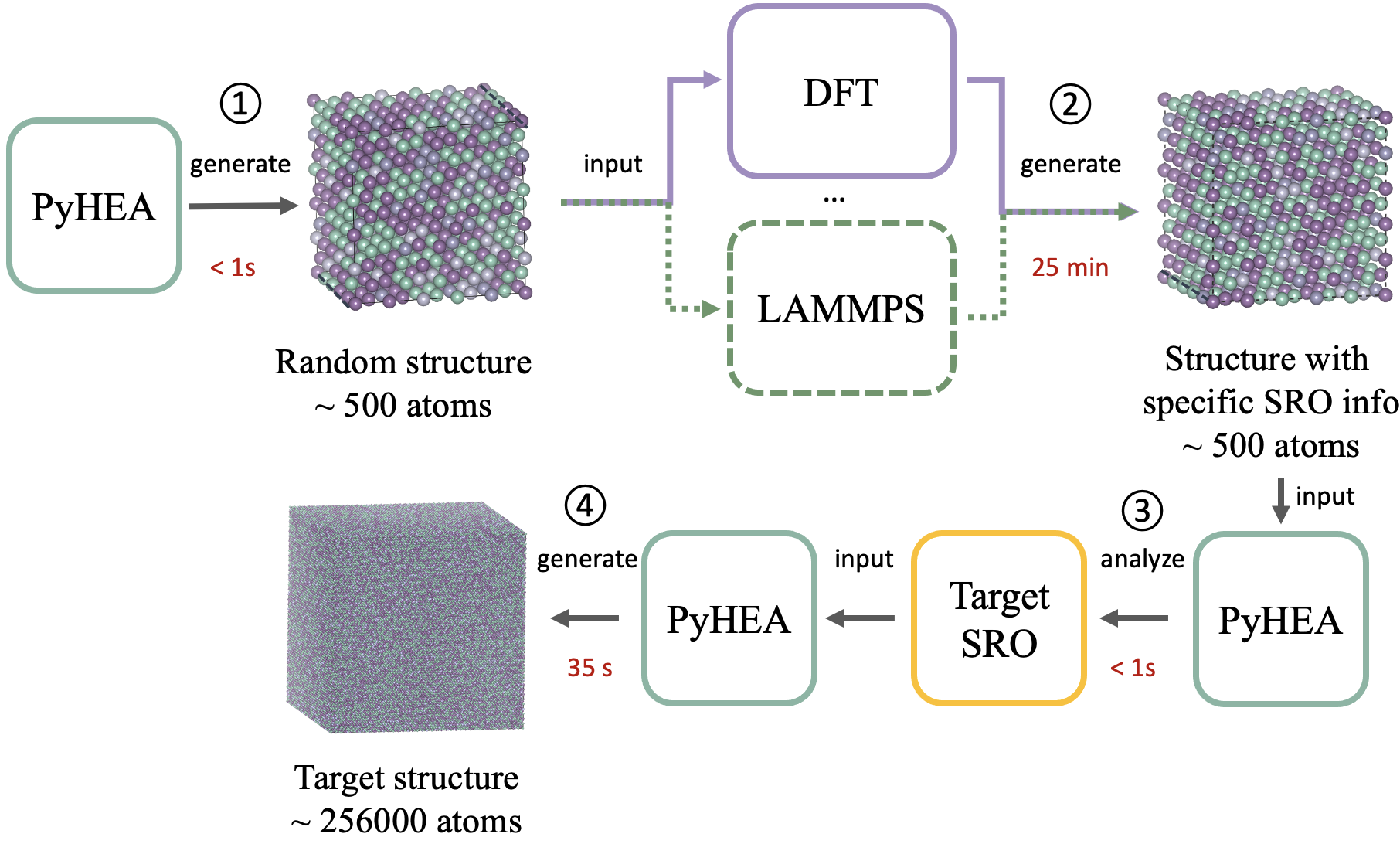}
    \caption{
    Schematic illustration of the application workflow, with red labels indicating approximate computation times:
    (1)~An initial random structure (containing $\sim500$ atoms) is generated in under 1\,s;
    (2)~this structure is passed to external tools (e.g., LAMMPS or first-principles calculations) for SRO evaluation (taking $\sim25$\,min in LAMMPS);
    (3)~the resulting structure is analyzed in under 1\,s to determine the target SRO;
    (4)~a large-scale target structure (containing $256{,}000$ atoms) is constructed in about 35\,s, based on the specified SRO parameters.
    }
    \label{fig:app_workflow}
\end{figure}

As illustrated in \autoref{fig:app_workflow}, PyHEA significantly streamlines the modeling workflow for HEAs. A small-scale system of approximately 500 atoms is first generated and refined (details of the Monte Carlo steps and convergence criteria are provided in \autoref{sec:methods}), after which the final SRO parameters are extracted. These SRO parameters guide the construction of a large-scale configuration containing up to $256{,}000$ atoms, a process completed in mere 35 seconds—substantially shorter than the hours or days typically required by Monte Carlo–Molecular Dynamics (MC–MD) simulations.

\begin{figure}[!htb]
    \centering
    \includegraphics[width=\textwidth]{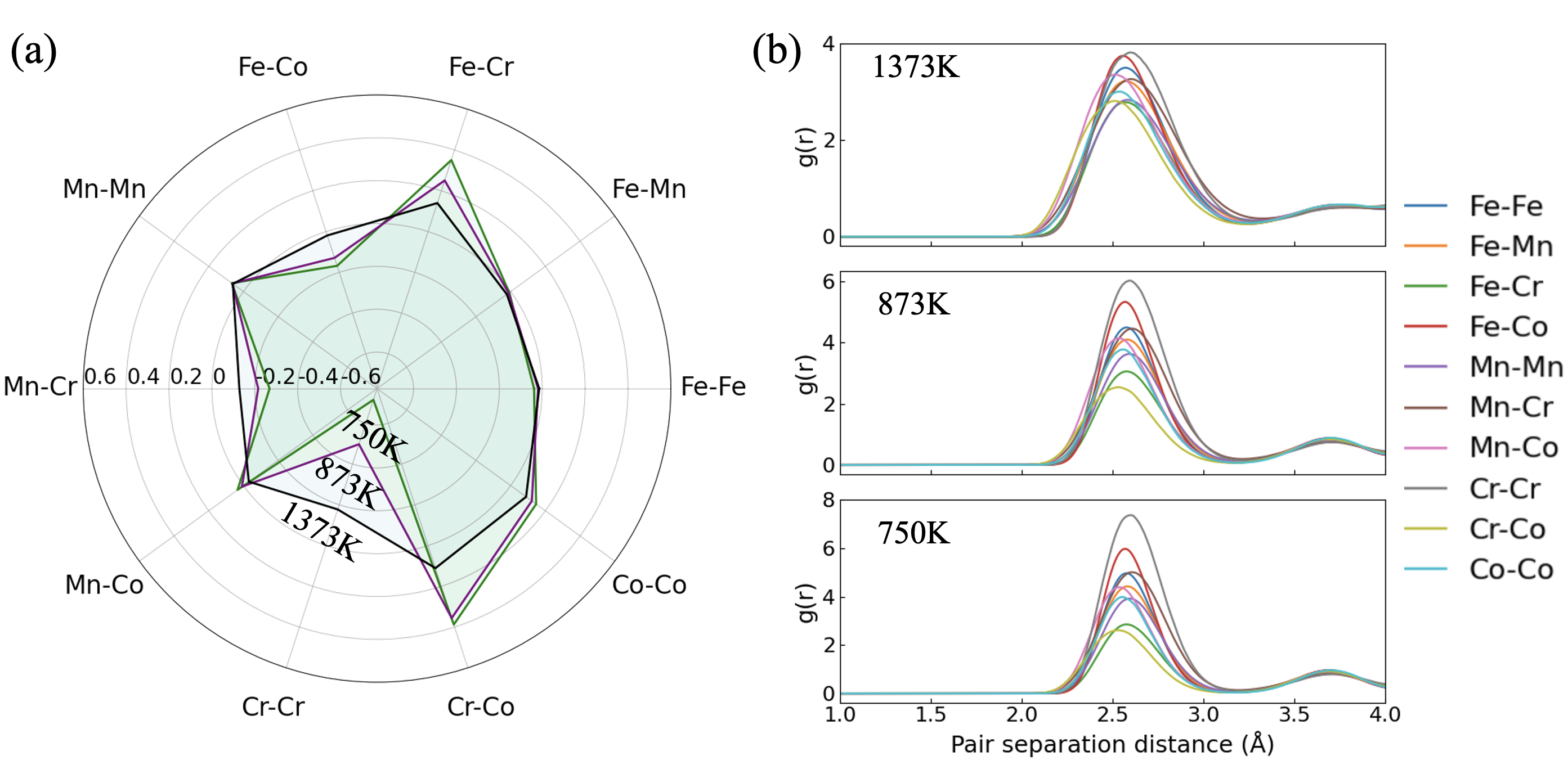}
    \caption{
    SRO values and Radial distribution function (RDF) for key atomic pairs at various temperatures in a $256{,}000$-atom Fe--Mn--Cr--Co system.
    (a)~A radar plot illustrating SRO values for selected atomic pairs (e.g., Fe--Fe, Fe--Mn, Fe--Cr) at three temperatures: 750\,K, 873\,K, and 1373\,K.
    (b)~RDF curves showing the density of atomic pairs as a function of separation distance (\AA) at the same temperatures.
    }
    \label{fig:app1}
\end{figure}

To demonstrate this workflow, we generated a $256{,}000$-atom Fe--Mn--Cr--Co (40\,\% Fe, 40\,\% Mn, 10\,\% Cr, 10\,\% Co) system using PyHEA, then performed a 10\,ps LAMMPS MD simulation in an NPT ensemble at multiple temperatures to obtain radial distribution function (RDF) data. \autoref{fig:app1} shows both the RDF and the SRO for representative atomic pairs. The RDF peaks indicate the positions of the first and second coordination shells, while the corresponding SRO values reveal how strongly specific atomic species tend to cluster or avoid one another.

The RDF and SRO curves exhibit a clear transition from a random-like distribution at 1373\,K to increasingly pronounced SRO configurations at 750\,K. This trend aligns qualitatively with prior computational and experimental studies on Fe--Mn--Cr--Co alloys, underscoring the physical reliability and robustness of our approach.

Finally, additional temperature-dependent SRO details are illustrated in \autoref{fig:apendix_heat_map}, which visualizes the Warren--Cowley SRO parameters at 1373\,K, 873\,K, and 750\,K. Taken together, these results underscore PyHEA’s efficiency in handling large-scale atomistic simulations and its capability to produce physically meaningful predictions that align well with established literature.

\section{Discussion}
\label{sec:discussion}

This study presents PyHEA as a computational framework that significantly advances HEA modeling by offering dramatic improvements in simulation speed without compromising accuracy. Compared to established tools such as ATAT-mcsqs and SCRAPs, PyHEA achieves speedups of up to 333,000× and 13,900×, respectively (\autoref{tab:performance}), while consistently demonstrating tighter fitness distributions (see \autoref{fig:compare}). These findings confirm that PyHEA maintains or even enhances precision across a wide range of system sizes.

A key factor in PyHEA’s scalability is its ability to handle systems with more than 256,000 atoms. Previous approaches often become intractable beyond a few thousand atoms, yet PyHEA’s GPU-accelerated design and incremental fitness calculations (detailed in \autoref{sec:incremental_fitness}) make large scale simulations both feasible and efficient. In the Fe--Mn--Cr--Co case study, for instance, PyHEA reduces computation time from several hours or days to minutes while reliably reproducing the temperature-dependent SRO patterns demonstrated in previous work.

To explain why PyHEA outperforms existing methods such as SCRAPs, it is crucial to highlight two technical innovations. First, the incremental SRO algorithm recalculates only the local environment affected by each atomic swap rather than recomputing the entire system’s SRO at every iteration. As described in Section~\ref{sec:incremental_fitness}, this approach lowers SRO-related overhead by up to 99.6\%. Second, PyHEA capitalizes on GPU acceleration to parallelize local Monte Carlo perturbations and incremental fitness evaluations, thereby multiplying the benefits of the incremental SRO strategy. In contrast, SCRAPs relies on a more conventional, system-wide recalculation of SRO and does not exploit large-scale GPU parallelization; hence it cannot match PyHEA’s order-of-magnitude speedups.

Beyond delivering efficiency and accuracy, PyHEA significantly accelerates the research cycle itself. By reducing simulation time from days to mere minutes or seconds, researchers can now quickly test hypotheses, tune parameters, and explore diverse atomic configurations that were previously too time-consuming. This capability supports a more iterative approach to materials design, enabling deeper and faster insights into composition--structure--property relationships in HEAs.

Moreover, PyHEA’s ability to simulate large systems with complex SRO phenomena under various temperatures and compositions provides valuable information for understanding structure-sensitive properties. In particular, it opens opportunities for investigating how SRO influences mechanical and thermal behaviors. The demonstrated consistency with both experimental and prior computational data reinforces the reliability of PyHEA’s predictions in practical contexts.

Looking ahead, PyHEA can be extended through integration with machine learning models, especially reinforcement learning, to guide or refine its global search. Incorporating more sophisticated interatomic potentials or linking to first-principles simulations would further broaden its applicability. Additional validation in collaboration with experimental teams would strengthen confidence in PyHEA’s predictive accuracy under different thermomechanical conditions.

In conclusion, PyHEA represents a substantial breakthrough in computational materials science by bridging the gap between theory and large-scale HEA simulations. Its use of incremental SRO calculations and GPU acceleration explains its marked speedups over methods like SCRAPs. By enabling full-atomistic simulations of tens or hundreds of thousands of atoms in a fraction of the time, PyHEA paves the way for faster, more exploratory HEA research and optimization, thus laying a foundation for future advancements in alloy development and discovery.

\section{Methods}
\label{sec:methods}

\begin{figure}[!htb]
    \centering
    \includegraphics[width=\textwidth]{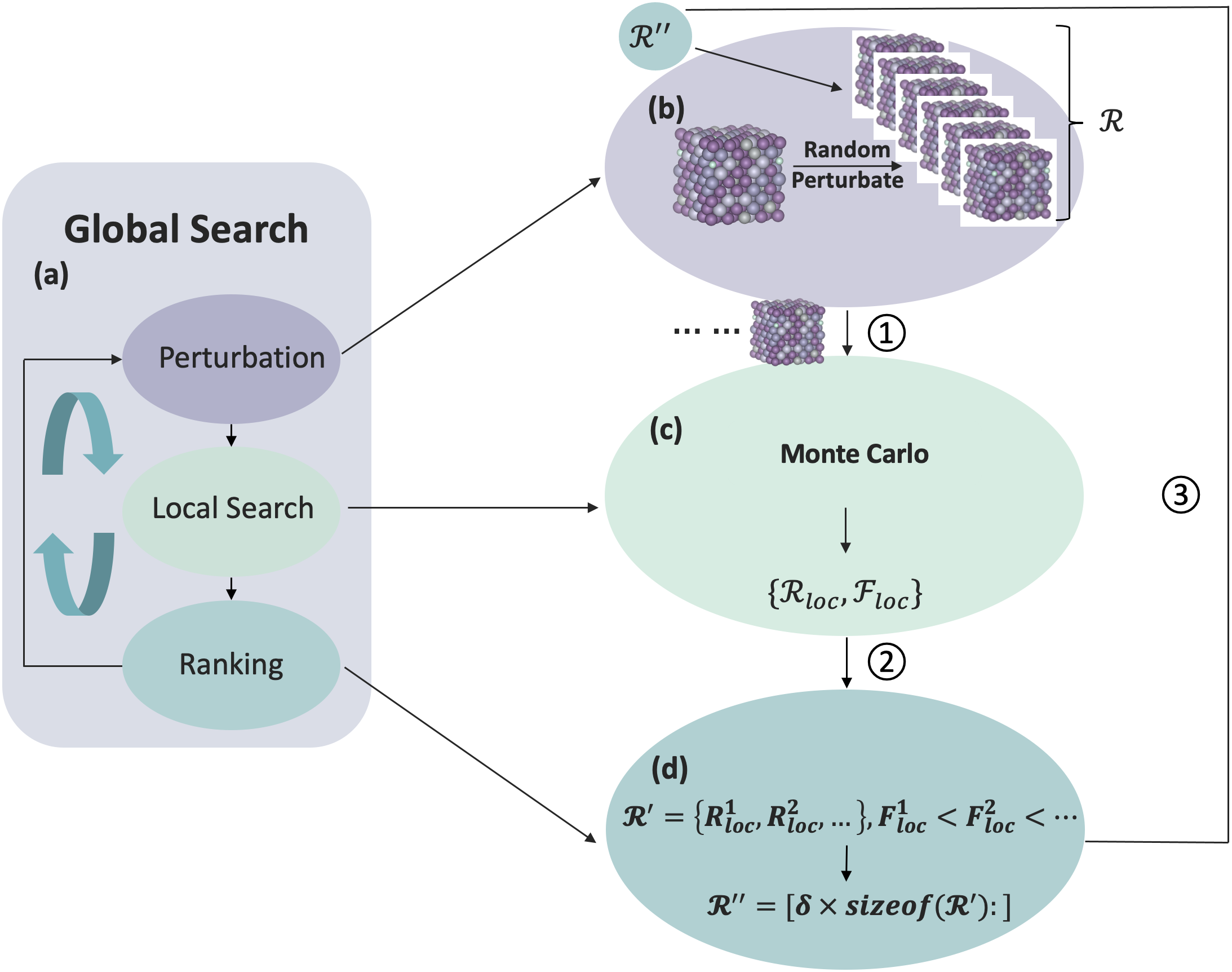}
    \caption{
    Workflow of the PyHEA algorithm for optimizing atomic configurations. 
    (a)~The global search framework iteratively cycles through the stages of perturbation, local search, and ranking to converge toward an optimal solution. 
    (b)~Initial atomic configurations consist of randomly generated and previously identified optimal structures for exploration. 
    (c)~Monte Carlo (MC) methods are applied to refine each candidate configuration and evaluate their fitness. 
    (d)~Configurations are ranked based on fitness, and the lowest-ranked fraction of solutions is discarded to enhance the selection process in subsequent iterations.
    }
    \label{fig:workflow}
\end{figure}

This section details the methods employed in our framework for optimizing atomic configurations in HEAs. We discuss the global search at \autoref{sec:global_search} for exploring the configuration space, the local search at \autoref{sec:mc} for refining promising candidates, SRO metrics at \autoref{sec:sro} for quantifying atomic arrangements, fitness calculation at \autoref{sec:fitness_calculation} for evaluating configurations, incremental updates at \autoref{sec:incremental_fitness} for enhancing computational efficiency, GPU implementation at \autoref{sec:gpu} for boosting the performance with heterogeneous computing devices, and finally MD simulations at \autoref{sec:md} for structure optimization and analysis.  \autoref{fig:workflow} provides an overview of the PyHEA workflow, summarizing its main components and their interactions.

\subsection{Global Search}
\label{sec:global_search}

\begin{algorithm}
    \caption{Global Search Algorithm}
    \label{alg:global_search}
    \begin{algorithmic}[1]
    \Require Input parameters, optimization function
    \Ensure Optimized solution(s)
    \State Initialize solutions (candidate configurations)
    \While{iteration $<$ Global maximum number}
        \State Create new solutions via random perturbations
        \State Calculate fitness $F$ for all solutions
        \State Select a fraction of solutions with the best fitness (top solutions)
        \For{each solution in top solutions}
            \State Refine solution via Monte Carlo (local search)
        \EndFor
        \State Discard the worst fraction $p_a$ of solutions
        \State Rank the solutions and identify the current best
    \EndWhile
    \State \Return best solutions
    \end{algorithmic}
\end{algorithm}

The optimization of atomic configurations in HEAs presents a complex challenge due to the vast solution space. As illustrated in \autoref{fig:workflow}(a), our global search framework implements an iterative optimization strategy that progressively refines atomic configurations through multiple cycles of exploration and evaluation, following the main workflow outlined in Algorithm~\ref{alg:global_search}. The effectiveness of this strategy is governed by several key parameters, including the number of candidate solutions and global iterations, which significantly influence the balance between exploration breadth and computational efficiency (as detailed in ~\autoref{sec:param_opt}).

The search process begins with the initialization of candidate solutions representing different atomic arrangements (\autoref{fig:workflow}(b)). The number of these initial solutions directly affects the search space coverage, with larger solution pools enabling more comprehensive exploration at the cost of increased computational overhead. During the first iteration, these configurations are generated entirely randomly to ensure comprehensive coverage of the solution space. In subsequent iterations, the solution pool combines newly generated random configurations with top-performing solutions retained from the previous iteration. Crucially, we maintain the number of retained solutions at less than or equal to half of the total configurations selected for local search, thereby ensuring sufficient randomness in each iteration while preventing premature convergence to local optima.

Each candidate solution undergoes fitness evaluation (\autoref{sec:fitness_calculation}) to quantify its alignment with target SRO characteristics. Following evaluation, a predetermined number of the highest-ranking configurations are selected for refinement through local search (\autoref{fig:workflow}(c)). The number of parallel tasks during this phase and the search depth of the Monte Carlo optimization significantly impact both the quality of refinement and computational efficiency. This local optimization phase, detailed in ~\autoref{sec:mc}, employs Monte Carlo perturbations to systematically explore the neighborhood of promising configurations, thereby improving their SRO properties.

The refined configurations then undergo a comprehensive ranking process, as illustrated by \autoref{fig:workflow}(d). At each iteration's conclusion, a fixed fraction $p_a$ of the lowest-performing solutions is eliminated, while the top-performing configurations are preserved as seeds for the subsequent iteration. This selection mechanism, combined with the continuous introduction of new random configurations, maintains an effective balance between preserving successful structural patterns and exploring new possibilities within the vast configuration space.

This iterative optimization process, as described in Algorithm~\ref{alg:global_search}, continues until reaching the maximum number of global iterations. The optimal number of iterations, as demonstrated in ~\autoref{sec:param_opt}, is crucial for achieving convergence while maintaining computational efficiency. Through careful parameter tuning, the algorithm progressively converges toward optimal atomic configurations that exhibit desired SRO characteristics.

\subsection{Local Search}
\label{sec:mc}

\begin{algorithm}
   \caption{Serial Monte Carlo Search Algorithm}
   \label{alg:serial_mc}
   \begin{algorithmic}[1]
   \Require Configuration $x$, optimization function $F$
   \Ensure Refined configuration
   \State Initialize current configuration $x_{\text{curr}} \leftarrow x$
   \While{iteration $<$ Local\_iterations}
       \State Generate a new solution $x_{\text{new}}$
       \State Calculate $\delta F = F(x_{\text{curr}}) - F(x_{\text{new}})$
       \If{$\delta F < 0$ \textbf{or} $|\delta F| \geq \text{threshold}$}
           \State Update $x_{\text{curr}} \leftarrow x_{\text{new}}$
       \EndIf
   \EndWhile
   \State \Return $x_{\text{curr}}$
   \end{algorithmic}
\end{algorithm}

\begin{algorithm}[htb]
   \caption{Parallel Monte Carlo Search Algorithm}
   \label{alg:parallel_mc}
   \begin{algorithmic}[1]
   \Require Configuration $x$, optimization function $F$
   \Ensure Refined configuration
   \State Initialize current configuration $x_{\text{curr}} \leftarrow x$
   \For{each task $i$ in parallel}
       \State Load $x_{\text{curr}}$ and $F(x_{\text{curr}})$
       \State Generate a new solution $x_i$
       \State Calculate $\delta F_i = F(x_{\text{curr}}) - F(x_i)$
       \State Synchronize tasks, then sort all $\delta F_i$ values
       \If{$\delta F_{\min} > 0$ \textbf{or} $|\delta F_{\min}| \leq \text{threshold}$}
           \State $depth \leftarrow 0$
           \State Update $x_{\text{curr}} \leftarrow x_i$
       \Else
           \State $depth \leftarrow depth + 1$
       \EndIf
       \If{$depth \geq \text{threshold}$}
           \State \textbf{break}
       \EndIf
   \EndFor
   \State \Return $x_{\text{curr}}$
   \end{algorithmic}
\end{algorithm}

While the global search framework effectively identifies promising configurations through iterative exploration, achieving optimal SRO characteristics requires more focused refinement. To address this need, we implement a specialized local search phase employing Monte Carlo techniques with both serial (Algorithm~\ref{alg:serial_mc}) and parallel (Algorithm~\ref{alg:parallel_mc}) approaches, with the latter specifically optimized for GPU acceleration to handle large-scale systems efficiently.

In the serial implementation, each configuration undergoes sequential refinement through iterative solution generation and evaluation. The algorithm computes the change in fitness ($\delta F$) for each new solution, accepting modifications that either improve the configuration ($\delta F < 0$) or meet specific threshold criteria. While this approach is effective for smaller systems, it becomes computationally intensive as system size increases, motivating our development of a parallel implementation.

The parallel Monte Carlo search (Algorithm~\ref{alg:parallel_mc}) significantly enhances efficiency by simultaneously evaluating multiple candidate solutions. For each solution, multiple tasks operate in parallel, where each task generates a new configuration by randomly swapping the positions of two atoms. As demonstrated in \autoref{sec:param_opt}, the number of parallel tasks represents a crucial parameter, with our analysis revealing optimal performance in the range of 64--128 tasks. After each round of atomic swaps, all tasks synchronize to evaluate the fitness changes and identify the most promising configurations. This coordinated approach ensures that computational resources are focused on the most promising regions of the solution space, while the depth-based termination criterion prevents excessive computation in regions unlikely to yield significant improvements.

A key innovation in our methods is the introduction of the search depth parameter, which serves as both a convergence criterion and an efficiency control mechanism. As shown in Algorithm~\ref{alg:parallel_mc}, the depth counter ($depth$) tracks consecutive non-improving iterations, incrementing when no better solutions are found and resetting to zero when improvements occur. Our parametric studies (\autoref{sec:param_opt}) indicate that a search depth threshold of 8--10 provides an optimal balance between solution quality and computational efficiency. When the depth counter reaches this threshold, indicating a sustained lack of improvement, the local search terminates and returns the best configuration found.

Another key innovation in our local search is the incremental fitness calculation. Instead of recalculating the entire system's SRO after each atomic swap, we update only the local regions affected by the swap operation. For large systems containing thousands of atoms, this approach reduces the computational overhead by up to 99.6\% compared to traditional full recalculation methods. The technical details and implementation of this incremental strategy are discussed thoroughly in \autoref{sec:incremental_fitness}.

\subsection{Short-Range Order}
\label{sec:sro}

Short-range order (SRO) describes how atoms in a multi-component system tend to cluster or repel each other relative to a random distribution \cite{zhang2017local}. We quantify SRO using the Warren--Cowley parameter \cite{cowley1950approximate}, denoted as $\alpha_{ij}^{(k)}$, which characterizes the degree of ordering between atomic species $i$ and $j$ in the $k$-th coordination shell:
\begin{equation} 
\label{eq:sro}
\alpha_{ij}^{(k)} = 1 - \frac{P_{ij}^{(k)}}{c_j},
\end{equation}
where $P_{ij}^{(k)}$ represents the conditional probability of finding a type-$j$ atom in the $k$-th shell around a type-$i$ atom, and $c_j$ is the overall concentration of type-$j$ atoms in the system. The parameter $\alpha_{ij}^{(k)}$ ranges from $-1$ to $+1$, where negative values indicate preferential attraction between species $i$ and $j$, positive values suggest repulsion, and zero corresponds to a random distribution.



\subsection{Fitness Calculation}
\label{sec:fitness_calculation}

To evaluate how well a given atomic configuration matches a target SRO pattern, we define a fitness function that incorporates the Warren--Cowley parameters:
\begin{align}
\label{eq:fitness_function}
F(X) &= \sqrt{
    \sum_{k=1}^{N_{\mathrm{shells}}} 
      w_k
      \sum_{i=1}^{N_{\mathrm{types}}}
      \sum_{j=1}^{N_{\mathrm{types}}}
        \bigl(
            \alpha_{ij}^{(k)}(X) 
            - 
            \alpha_{ij}^{(k),\mathrm{target}}
        \bigr)^2
}\,, 
\end{align}
where $k$ indexes the coordination shells ($1 \leq k \leq N_{\mathrm{shells}}$), and $i,j$ enumerate the atomic species ($1 \leq i,j \leq N_{\mathrm{types}}$). Here, $\alpha_{ij}^{(k)}(X)$ is computed according to Eq.(1) for configuration $X$. The shell-specific weights $w_k$ and target SRO values $\alpha_{ij}^{(k),\mathrm{target}}$ are user-specified input parameters, allowing flexible control over the optimization process. The weights $w_k$ can be adjusted to emphasize certain coordination shells, while $\alpha_{ij}^{(k),\mathrm{target}}$ values can be derived from experimental measurements, theoretical calculations, or other simulation results to guide the structure generation toward desired atomic ordering patterns.

\subsection{Incremental Calculation of Fitness}
\label{sec:incremental_fitness}

\begin{figure}[!htp]
    \centering
    \includegraphics[width=\columnwidth]{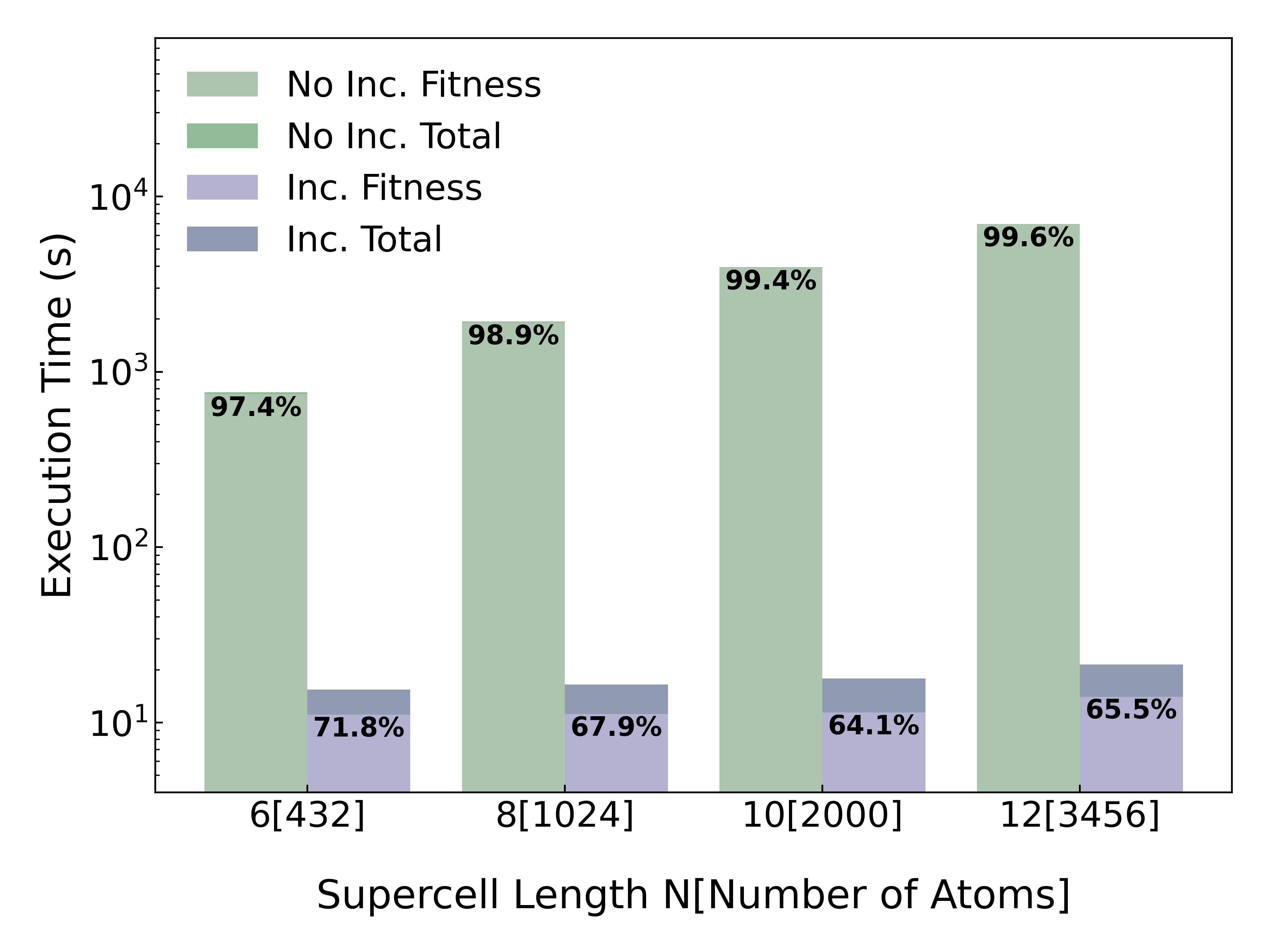}
    \caption{
    Comparison of execution times for incremental and non-incremental fitness and total calculations in Monte Carlo search. The bar chart shows the execution times (in seconds) for different supercell lengths N and corresponding atom counts. Incremental methods demonstrate substantial improvements, reducing execution time for both fitness evaluation and total computation. The incremental approach achieves up to a 99.6\% reduction in total computation time for larger supercells.
    }
    \label{fig:sro_benchmark}
\end{figure}

In a non-incremental approach, the Warren--Cowley parameters and resulting fitness are recomputed from scratch at each Monte Carlo step (\autoref{eq:sro} and \autoref{eq:fitness_function}), which becomes computationally prohibitive for large systems. As shown in \autoref{fig:sro_benchmark}, for supercells containing thousands of atoms, these calculations can consume over 99\% of the total execution time.

To address this bottleneck, we implement an incremental fitness update strategy. Rather than recalculating all $\alpha_{ij}^{(k)}$ values after each atomic swap, we update only those parameters affected by the exchanged atoms $i$ and $j$. The change in the Warren--Cowley parameter for the $k$-th shell is given by:
\begin{equation} 
\Delta \alpha_{ij}^{(k)} = \alpha_{ij}^{(k),\mathrm{new}} - \alpha_{ij}^{(k),\mathrm{old}} = 
\left(1 - \frac{P_{ij}^{(k),\mathrm{new}}}{c_j}\right) -
\left(1 - \frac{P_{ij}^{(k),\mathrm{old}}}{c_j}\right),
\end{equation}
where $P_{ij}^{(k),\mathrm{new}}$ and $P_{ij}^{(k),\mathrm{old}}$ are the probabilities before and after the swap. Since only atoms within the local environment of sites $i$ and $j$ are affected, this reduces the computational complexity from $O(N)$ to $O(1)$ per Monte Carlo step.

\subsection{GPU Optimization}
\label{sec:gpu}
To maximize computational efficiency for large-scale systems, we implement a parallel GPU-accelerated version of the Monte Carlo search. The implementation follows a hierarchical parallelization strategy where each GPU thread processes an individual atomic swap operation, while thread blocks manage complete Monte Carlo refinements for distinct candidate solutions. This design enables concurrent optimization of multiple configurations while efficiently utilizing GPU resources including registers, shared memory, and mixed-precision arithmetic.
\subsubsection{Parallel Processing Architecture}
The GPU implementation distributes computation across multiple levels:
\begin{itemize}
\item \textbf{Thread-Level Parallelism:} Individual threads handle SRO calculations for specific atomic pairs using reduced precision (FP16, Uint16) data types, balancing numerical accuracy with memory efficiency.
\item \textbf{Block-Level Management:} Each thread block maintains a candidate configuration in shared memory, following the lattice structure described in \cite{dahlborg2016structure}. This arrangement minimizes global memory access during atomic position updates.
\end{itemize}

\subsubsection{Incremental Fitness Evaluation}

Following the incremental update strategy detailed in \autoref{sec:incremental_fitness}, each thread computes local fitness changes according to:

\begin{equation}
F_{\text{new}} = F_{\text{prev}} + \Delta F_{\text{perturb}},
\end{equation}

where $\Delta F_{\text{perturb}}$ is calculated from the changes in Warren--Cowley parameters ($\Delta \alpha_{ij}^{(k)}$) affected by the atomic swap. This approach maintains consistency with the fitness function defined in \autoref{eq:fitness_function} while exploiting the locality of updates.

\subsubsection{Parallel Selection Strategy}

After computing fitness changes, threads within each block cooperatively sort the results using shared memory operations. The best configuration update is identified according to:

\begin{equation}
F_{\text{best}} = \min_{\text{updates}} {F_{\text{new}}},
\end{equation}

ensuring that each block maintains its optimal configuration throughout the refinement process.

\subsubsection{Convergence Criteria}

The GPU-accelerated search terminates when either:
\begin{itemize}
\item The fitness value $F_{\text{best}}$ remains unchanged for a number of iterations exceeding the search depth threshold defined in \autoref{sec:mc}, or
\item The maximum number of Monte Carlo steps is reached.
\end{itemize}

Upon convergence, the final atomic configuration and its associated fitness value are transferred from GPU to CPU memory for subsequent analysis or further refinement cycles.
This GPU-optimized implementation achieves significant speedups over CPU-based calculations (see \autoref{tab:performance}), enabling the efficient handling of systems containing hundreds of thousands of atoms while maintaining the accuracy of SRO optimization. The approach seamlessly integrates with the broader optimization framework described in \autoref{sec:methods}, providing a scalable solution for large-scale HEA modeling.

\subsection{Molecular Dynamics Simulations}
\label{sec:md}
Molecular dynamics (MD) simulations were performed using LAMMPS \cite{li2016mechanical, afkham2017tensile, tang2021nano} for two distinct purposes: generating configurations enriched with SRO features and validating the large-scale structures generated by PyHEA.

To generate configurations with target SRO properties, MD simulations were conducted on a 500-atom system under specific temperature and pressure conditions. An NPT ensemble with a timestep of 2\,fs was used, and the simulation ran for 200\,ps. To accelerate SRO formation, an MC-MD hybrid method was applied, involving random atom swaps every 0.2\,ps. The final configuration was extracted for subsequent analysis, ensuring alignment with the desired SRO characteristics.

For validation, MD simulations were carried out on PyHEA-generated configurations containing up to 256,000 atoms. These simulations, also performed in an NPT ensemble with a 2\,fs timestep, ran for 10\,ps. RDFs were calculated to evaluate the structural reliability of the large-scale configurations, comparing them against known computational and experimental benchmarks.These complementary simulations ensured that SRO structures were accurately generated in small systems and validated in large-scale configurations, providing a robust framework for modeling high-entropy alloys.

Overall, these integrated computational techniques—global and local searches, incremental SRO updates, GPU acceleration, and complementary MD—form a robust toolkit for generating and analyzing large-scale HEA models efficiently and accurately.

\section{Acknowledgment}

This work was conducted under the supervision of Professor Lifeng Liu, School of Integrated Circuits, Peking University. The author acknowledges the financial support provided by the National Natural Science Foundation of China (Grant No. 61874006).

\section{Code Availability}

The code used for this study is available at \url{https://github.com/caimeiniu/pyhea}. Access is granted under LGPL-3.0 license.


\appendix

\section{Supplementary Data}

\begin{table}[H]
    \centering
    \begin{threeparttable}
    \caption{Performance comparison between PyHEA and existing methods under parallel computing conditions. Both PyHEA and SCRAPs utilized 24 CPU cores (Intel Xeon Gold 6132 \text{@} 2.60GHz) for parallel execution, with additional GPU (single NVIDIA 4090) results shown for PyHEA. The table presents computation times (in minutes) and speedup ratios for both body-centered cubic (bcc) and face-centered cubic (fcc) structures of varying sizes (\#At) and species numbers (\#Sp). All simulations maintained consistent parameters: 128 candidate solutions, 10 global iterations, 256 MC tasks, a search depth of 10, and supercell sizes following $S = A \cdot L^3$ where $A = 2$ (bcc) or $A = 4$ (fcc). SRO was optimized over 3 coordination shells. Dashes (--) indicate computationally infeasible cases, while speedup values are shown relative to SCRAPs[ATAT-mcsqs in brackets].}
    \label{tab:parallel}
    \begin{tabular}{c|cc|c|cc|c|cc}
        \Xhline{0.7pt}
        \multirow{2}{*}{Type} & \multirow{2}{*}{\#Sp\tnote{a}} & \multirow{2}{*}{\#At\tnote{b}} & ATAT- & SCRAPs & PyHEA & PyHEA & \multicolumn{2}{c}{Speedup\tnote{c}} \\
        & & & mcsqs & Parallel & Parallel & GPU & CPU & GPU \\
        \Xhline{0.7pt}
        bcc & 3 & 54 & \textgreater 1.4k\tnote{d} & 0.14 & 0.14 & .03 & 1.0[10k] & 4.67[47k]\\
        bcc & 4 & 128 & \textgreater 10k\tnote{d} & 0.34 & 0.19 & .03 & 1.79[53k] & 11.33[333k]\\
        bcc & 4 & 432 & -- & 1.26 & 0.28 & .03 & 4.5 & 42\\
        bcc & 4 & 1024 & -- & 3.18 & 0.32 & .03 & 9.94 & 106\\
        bcc & 4 & 2048 & -- & 7.02 & 0.36 & .04 & 19.5 & 175.5\\
        bcc & 5 & 2048 & -- & 7.27 & 0.48 & .04 & 15.2 & 181.8\\
        bcc & 5 & 6.8k & -- & 47.4 & 0.71 & .05 & 66.8 & 948\\
        bcc & 6 & 28k & -- & -- & 2.05 & .09 & -- & --\\
        \hline
        fcc & 3 & 108 & -- & 0.63 & 0.17 & .01 & 3.71 & 63\\
        fcc & 4 & 256 & -- & 2.24 & 0.27 & .01 & 8.3 & 224\\
        fcc & 5 & 500 & -- & 5.81 & 0.43 & .01 & 13.5 & 581\\
        fcc & 5 & 4k & -- & 64.3 & 0.65 & .01 & 99.0 & 6.4k\\
        fcc & 4 & 256k & -- & -- & 61.1 & .58 & -- & --\\
        \Xhline{0.7pt}
    \end{tabular}
    \begin{tablenotes}
    \small
        \item[a] Notation: \#Sp = number of elemental species.
        \item[b] Notation: \#At = total number of atoms in the system.
        \item[c] Speedup ratios: value vs SCRAPs [value vs ATAT-mcsqs in brackets].
        \item[d] ATAT-mcsqs data from Ref. \cite{singh2021accelerating}. All times in minutes.
    \end{tablenotes}
    \end{threeparttable}
\end{table}

\begin{figure}[H]
    \centering
    \includegraphics[width=\textwidth]{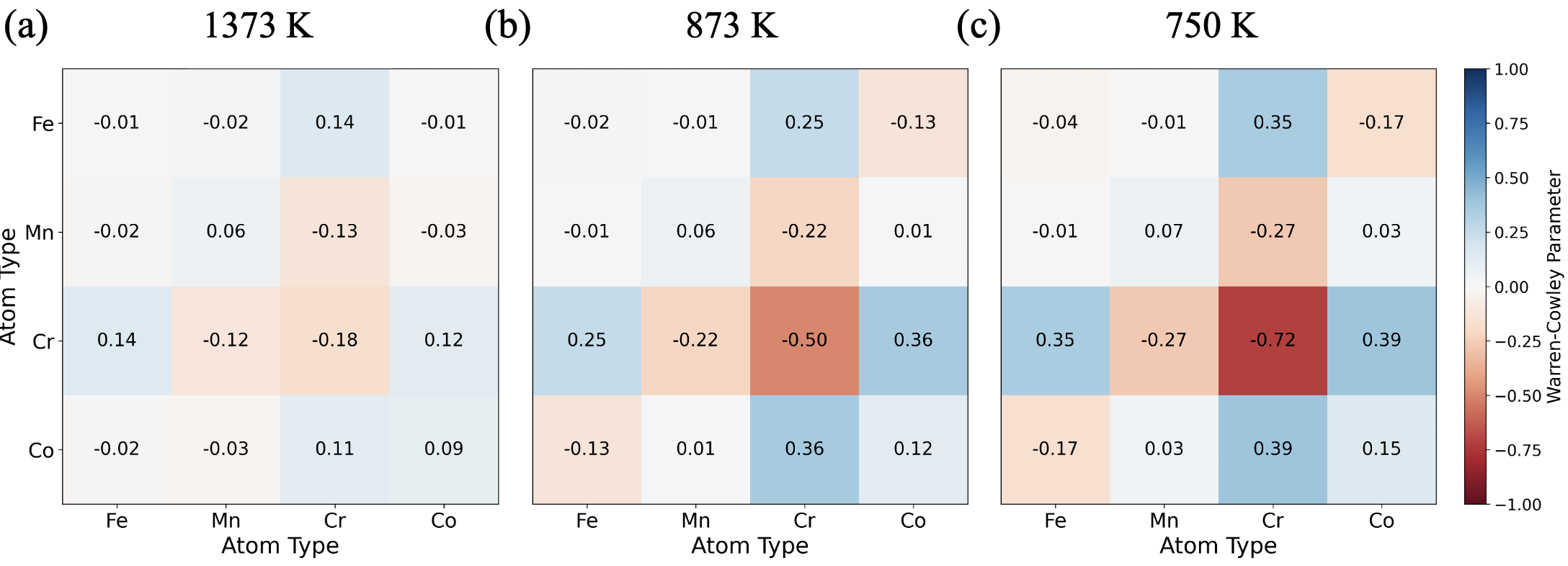}
    \caption{
    Warren–Cowley SRO parameter matrices for a multi-principal-element alloy (Fe, Mn, Cr, Co) at three different temperatures: (a) 1373 K, (b) 873 K, and (c) 873 K. The color scale ranges from blue (positive values) to red (negative values), indicating a preference for certain atom types to cluster together (positive SRO) or to avoid each other (negative SRO). At high temperatures (1373 K), the SRO values are near zero, reflecting a more random distribution, whereas at lower temperatures (873 K), the emergence of pronounced SRO patterns reveals stronger local atomic ordering.
    }
    \label{fig:apendix_heat_map}
\end{figure}






\end{document}